\documentclass[twocolumn,prd,superscriptaddress,nofootinbib,showpacs]
{revtex4}

\usepackage{graphicx}
\usepackage{dcolumn}
\usepackage{bm}

\begin{document}

\title{A model for time-dependent cosmological constant and its
consistency with the present Friedmann universe}

\author{M. Novello}
\email{novello@cbpf.br}
\affiliation{Centro Brasileiro de Pesquisas F\'{\i}sicas,\\
Rua Dr.\ Xavier Sigaud 150, Urca 22290-180 Rio de Janeiro, RJ --
Brazil}
\author{J. Barcelos-Neto}
\email{barcelos@if.ufrj.br}
\affiliation{Instituto de F\'{\i}sica\\
Universidade Federal do Rio de Janeiro, RJ 21945-970 -- Brazil}
\author{J.M. Salim}
\email{jsalim@cbpf.br}
\affiliation{Centro Brasileiro de Pesquisas F\'{\i}sicas,\\
Rua Dr.\ Xavier Sigaud 150, Urca 22290-180 Rio de Janeiro, RJ --
Brazil}

\date{\today}

\begin{abstract}
We use a model where the cosmological term can be related to the
chiral gauge anomaly of a possible quantum scenario of the initial
evolution of the universe. We show that this term is compatible with
the Friedmann behavior of the present universe.
\end{abstract}


\maketitle

\section{Introduction}
The idea of a varying cosmological constant is not new but has been
systematically examined in the last two decades \cite{Overduin}. In
particular, with the advent of the inflation mechanism and with the
difficulties related to the graceful exit problem \cite{Kaloper},
various authors were forced to consider $\Lambda$ as ground state of a
scalar theory \cite{Ellis}. However, we may say that the first
concrete mechanism for a varying $\Lambda$ was proposed by Dolgov
\cite{Dolgov} based on a nominimally coupled scalar field. A sequence
of other papers come afterwards, but all of them also based on scalar
fields~\cite{Ozer}.

\medskip
In a previous work \cite{Novello} we have presented a different model
for a time dependent cosmological ``constant" based on gauge fields
where its origin was related to a possible quantum scenario of the
initial evolution of the universe. This was achieved by showing that
the chiral gauge anomaly could be conveniently adapted in order to
generate a cosmological {\it constant}. In this way, the presence of
this term today would be a reminiscence of that initial quantum
behavior of the universe.

\medskip
In the present paper, we are going to consider a particular scenario
where the geometry is initially Bianchi-like (spatially homogeneous
but anisotropic) \cite{Belinsky}. We show that the evolution obtained
from the Einstein equations in the presence of the cosmological term
(as well as the Maxwell one) leads to an asymptotic solution that is
compatible with the Friedmann universe \cite{Weinberg,Novello2}.

\medskip
Our paper is organized as follow. In Sec. II we review the general
features of the model. The application is done in Sec. III. In Sec. IV
we deal with the solution of the particular Einstein equations. We
left Sec. V for some concluding remarks and introduce an Appendix to
show some details of the calculations.

\section{Review of the model}
\renewcommand{\theequation}{2.\arabic{equation}}
\setcounter{equation}{0}

Let us consider an action with the following general form
\cite{Novello}

\begin{equation}
S_\Lambda=\int d^4x\,\sqrt{-g}\,\,Y(\cal G)
\label{2.1}
\end{equation}

\noindent
where $g$ is the determinant of the metric tensor and $Y$ is some
function of an invariant quantity $\cal G$ which is constructed in
terms of a gauge field strength $G_{\mu\nu}^{\rm a}$  and its dual
$^\ast G_{\mu\nu}^{\rm a}=\frac{1}{2}\,\eta_{\mu\nu\rho\lambda}
\,G^{{\rm a}\rho\lambda}$ as

\begin{equation}
{\cal G}=^\ast G^{a\mu\nu}G^a_{\mu\nu}
\label{2.1a}
\end{equation}

\noindent
We are using the following definition for $\eta_{\mu\nu\rho\lambda}$

\begin{equation}
\eta_{\mu\nu\rho\lambda}=\sqrt{-g}\,\epsilon_{\mu\nu\rho\lambda}
\label{2.2}
\end{equation}

\noindent
Consequently,

\begin{equation}
\eta^{\mu\nu\rho\lambda}=-\frac{1}{\sqrt{-g}}
\,\epsilon^{\mu\nu\rho\lambda}
\label{2.3}
\end{equation}

\noindent
where $\epsilon_{\mu\nu\rho\lambda}$ and
$\epsilon^{\mu\nu\rho\lambda}$ are the usual Levi-Civita tensor
densities ($\epsilon_{0123}=1$).

\medskip
The variation of $S_\Lambda$ with respect the metric tensor leads to

\begin{eqnarray}
\frac{\delta S_\Lambda}{\delta g^{\mu\nu}}
&=&\int d^4x\,\biggl(\frac{\delta\sqrt{-g}}{\delta g^{\mu\nu}}\,Y
+\sqrt{-g}\,\frac{dY}{d{\cal G}}\,
\frac{\delta{\cal G}}{\delta g^{\mu\nu}}\biggr)
\nonumber\\
&=&\int d^4x\biggl(-\frac{1}{2}\,\sqrt{-g}\,g_{\mu\nu}\,Y
+\sqrt{-g}\,\frac{dY}{d{\cal G}}\,
\frac{1}{2}{\cal G}\,g_{\mu\nu}\biggr)
\nonumber\\
&=&\frac{1}{2}\int d^4x\,\sqrt{-g}\,
\Bigl(\frac{dY}{d{\cal G}}\,{\cal G}-Y\Bigr)\,g_{\mu\nu}
\label{2.4}
\end{eqnarray}

\noindent
We then observe that the action $S_\Lambda$ contributes to the
energy-momentum tensor with a term that is proportional to the metric
tensor. This is interpreted in the Einstein General Relativity theory
as a spacetime dependent cosmological {\it constant},

\begin{equation}
\Lambda=\frac{dY}{d{\cal G}}\,{\cal G}-Y
\label{2.5}
\end{equation}

\medskip
For a question of simplicity, we shall consider the field strength of
the Maxwell electromagnetic theory, where there is a natural
realization of this model coming from the chiral gauge anomaly.
For instance, from the path integral formalism of quantum matter and
gauge fields in a classical curved background one obtains to the
following effective action\cite{Birrell,Novello3}

\begin{eqnarray}
S_{\rm eff}&=&\frac{\theta}{4}\int d^4x\,\sqrt{-g}\,\alpha(x)\,
\eta^{\mu\nu\rho\lambda}F_{\mu\nu}F_{\rho\lambda}
\nonumber\\
&=&\frac{\theta}{4}\int d^4x\,\sqrt{-g}\,\alpha(x)\,{\cal G}
\label{2.6}
\end{eqnarray}

\noindent
where $\theta$ is a constant and $\alpha(x)$ is the gauge parameter
related to the chiral gauge transformation.

\medskip
The manner in which the effective action above is presented it is not
appopriated to be directly related to $S_\Lambda$. This is so because
since it is linear in $\cal G$ and cannot generate a cosmological term
$\Lambda$, as can be verified in Eq. (\ref{2.5}). However, this
problem can be circumvented by conveniently taking the generic
function $\alpha(x)$ of the action (\ref{2.6}) as ${\cal G}^p$, where
$p$ is, at first, any rational number~\empty
\footnote{The way of circumventing this problem here is slightly
different of the original paper \cite{Novello}. There, we have
redefined the gauge field in order to incorporate a nonlinearity of
$\cal G$.}.
In this way, the cosmological action $S_\Lambda$ turns to be

\begin{equation}
S_\Lambda=\frac{\theta}{4}\int d^4x\,\sqrt{-g}\,{\cal G}\,^{p+1}
\label{2.7}
\end{equation}

\noindent
We observe that for $p=-1$ we have an actual cosmological constant in
the Einstein equation.

\medskip
The next natural step is to consider this idea in some cosmological
model in order to see the way that the cosmological term, coming from
Eq. (\ref{2.7}), modifies the dynamics of the Einstein equation.

\section{Application of the model}
\renewcommand{\theequation}{3.\arabic{equation}}
\setcounter{equation}{0}

Let us start from the general action

\begin{eqnarray}
&&S=\int d^4x\,\sqrt{-g}\,\Bigl[\frac{1}{2\kappa}R
-\frac{1}{4}F_{\mu\nu}F^{\mu\nu}
\nonumber\\
&&\phantom{S=\int d^4x\,\sqrt{-g}\,\Bigl[\kappa R}
+\frac{1}{4}\theta(F_{\mu\nu}F^{\ast\mu\nu})^{p+1}
\nonumber\\
&&\phantom{S=\int d^4x\,\sqrt{-g}\,\Bigl[\kappa R}
+i\bar\psi\gamma^\mu(\nabla_\mu-ieA_\mu)\psi\Bigr]
\label{3.1}
\end{eqnarray}

\noindent
The equations of motion $\delta S/\delta g^{\mu\nu}=0$, $\delta
S/\delta A^\mu=0$, and $\delta S/\delta\psi=0$ lead respectively to

\begin{eqnarray}
&&G_{\mu\nu}=-F_{\mu\alpha}F^\alpha\,_\nu
-\frac{1}{4}\,g_{\mu\nu}\,F_{\alpha\beta}F^{\alpha\beta}
\nonumber\\
&&\phantom{G_{\mu\nu}=}
-\frac{p\theta}{2}g_{\mu\nu}(F_{\alpha\beta}F^{\ast\alpha\beta})^{p+1}
+T_{\mu\nu}^\psi
\label{3.2}\\
&&F^{\mu\nu}_{\phantom{\mu\nu};\nu}
-(p+1)\theta\bigl[(F_{\alpha\beta}F^{\ast\alpha\beta})^p
\bigr]_{,\nu}\,F^{\ast\mu\nu}
\nonumber\\
&&\phantom{G_{\mu\nu}}
=e\,\bar\psi\gamma^\mu\psi
\label{3.3}\\
&&\gamma^\mu\bigl(\nabla_\mu-ieA_\mu\bigr)\,\psi=0
\label{3.4}
\end{eqnarray}

\noindent
where $G_{\mu\nu}\equiv R_{\mu\nu}-\frac{1}{2}g_{\mu\nu}R$ is the
Einstein tensor and the constant $\kappa$ of (\ref{3.1}) was taken
$1$. In the obtainment of Eq. (\ref{3.3}) it was used the (Bianchi)
identity

\begin{equation}
F^{\ast\mu\nu}_{\phantom{\ast\mu\nu};\nu}=0
\label{3.5}
\end{equation}

We shall consider that $F^{\mu\nu}=F^{\mu\nu}(t)$. So, taking $\nu=0$
in Eq. (3.3), as well as $\mu=0$, we get

\begin{eqnarray}
&&\bar\psi\gamma^0\psi=\psi^\dagger\psi=0
\nonumber\\
&\Rightarrow&\psi=0
\label{3.6}
\end{eqnarray}

\noindent
From Eq. (\ref{3.5}) we also have that only $F^{\ast i0}$ can be
obtained. Using the Bianchi-like metric

\begin{equation}
ds^2=dt^2-a^2(t)dx^2-b^2(t)(dy^2+dz^2)
\label{3.7}
\end{equation}

\noindent
and just taking that $F^{\ast10}\neq0$, we have

\begin{eqnarray}
&&F^{\ast01}_{\phantom{\ast01};0}=0
\nonumber\\
&\Rightarrow&\sqrt{-g}\,F^{\ast01}_{\phantom{\ast01};0}=0
\nonumber\\
&\Rightarrow&\bigl(\sqrt{-g}\,F^{\ast01}\bigr)_{,0}=0
\nonumber\\
&\Rightarrow&\sqrt{-g}\,F^{\ast01}=constant
\nonumber\\
&\Rightarrow&F^{\ast01}=\frac{B_0}{ab^2}
\label{3.8}
\end{eqnarray}

\noindent
where, in the last step, the {\it constant} was identified as $B_0$
and we have used the metric given by (\ref{3.7}). From this result and
using Eq. (\ref{2.3}) we directly infer that

\begin{equation}
F_{23}=-B_0
\label{3.9}
\end{equation}

\medskip
However, the solution for $F^{10}$ depends on $p$. In fact, taking
$\nu=0$ and $\mu=1$ in Eq. (\ref{3.3}) and using the results given by
(\ref{3.6}) and (\ref{3.8}), we obtain

\begin{equation}
ab^2\,F^{10}+(p+1)\theta\,B_0^{p+1}\,
\Bigl(\frac{2aF^{10}}{b^2}\Bigr)^p=E_0
\label{3.10}
\end{equation}

\noindent
where $E_0$ was chosen to identify the constant that appears in the
solution of the corresponding differential equation.

\medskip
Our goal from now on is to look for if there exists some value of $p$
that renders a consistent solution for the Einstein equations
(\ref{3.2}) having in mind the Friedmann behavior of the present
universe. This will be the subject of next section. We conclude the
present section by emphasizing the importance of the Maxwell term,
$-\frac{1}{4}F_{\mu\nu}F^{\mu\nu}$, into the initial action
(\ref{3.1}). Without it, the only possible solution of (\ref{3.3})
would just be an actually constant. Indeed, taking $\psi=0$ and $\nu=
0$ in Eq. (\ref{3.3}), we get

\begin{equation}
F^{i0}_{\phantom{i0};0}
-(p+1)\theta\bigl[(F_{\alpha\beta}F^{\ast\alpha\beta})^p\bigr]_{,0}\,
F^{\ast i0}=0
\label{3.11}
\end{equation}

\noindent
where the first term of the expression above comes from the Maxwell
term. We then observe that without it we would obtain that
$(F_{\alpha\beta}F^{\ast\alpha\beta})^p$ should be a constant for
$p\neq-1$ [for $p=-1$, the cosmological term is also a constant
according to(\ref{2.7})].

\section{Particular solution of the Einstein equation}
\renewcommand{\theequation}{4.\arabic{equation}}
\setcounter{equation}{0}

\medskip
We have seen that to avoid an actual cosmological constant and the
trivial case without any cosmological term, $p$ cannot have the values
zero and minus one, respectively (and the Maxwell term has to be
present in the Lagrangian). In order to have a general view of the
physical behavior of the Einstein solution with the value of $p$, let
us initially choose $p=1$. This corresponds to one of the simplest
solution of Eq. (\ref{3.10}), that is

\begin{equation}
F^{10}=\frac{E_0b^2}{a(b^4+4\theta\,B_0^2)}
\label{4.1}
\end{equation}

\noindent
Combining this result with the ones given by (\ref{3.6}), (\ref{3.8}),
and (\ref{3.9}), we obtain that the Einstein equation (\ref{3.2})
leads to

\begin{eqnarray}
2\,\frac{\dot a\dot b}{ab}
+\Bigl(\frac{\dot b}{b}\Bigr)^2
&=&-\,\frac{1}{2}\Bigl[\frac{B_0^2}{b^4}
+\frac{E_0^2b^4}{(b^4+2\theta B_0)^2}\Bigr]
\nonumber\\
&&+\frac{2\theta E_0^2B_0^2}{(b^4+2\theta B_0)^2}
\label{4.2}\\
2\,\frac{\ddot b}{b}
+\Bigl(\frac{\dot b}{b}\Bigr)^2
&=&-\,\frac{1}{2}\Bigl[\frac{B_0^2}{b^4}
+\frac{E_0^2b^4}{(b^4+2\theta B_0)^2}\Bigr]
\nonumber\\
&&-\frac{2\theta E_0^2B_0^2}{(b^4+2\theta B_0)^2}
\label{4.3}\\
\frac{\ddot b}{b}+\frac{\ddot a}{a}+\frac{\dot a\dot b}{ab}
&=&\frac{1}{2}\Bigl[\frac{B_0^2}{b^4}
+\frac{E_0^2b^4}{(b^4+2\theta B_0)^2}\Bigr]
\nonumber\\
&&-\frac{2\theta E_0^2B_0^2}{(b^4+2\theta B_0)^2}
\label{4.4}
\end{eqnarray}

\noindent
We observe that the term related to the cosmological {\it constant}
has a behavior of $b^{-8}$ when $b$ goes to infinity (the behavior of
the radiation term is with $b^{-4}$). Consequently, the cosmological
term should disappear before the radiation era, what is not consistent
with the Friedmann universe, where $a=b$ (notice that it is the
Maxwell term which does not permit to have $a=b$ in the last two
equations).

\medskip
From the above results, it is not difficult to conclude that possible
values of $p$ that should be compatible with this asymptotic behavior
must stay between zero and minus one. However, for these values of
$p$, the solution of Eq. (\ref{3.10}) is not so direct as in the
previous case. For example, taking $p=-1/2$, where Eq. (\ref{3.10})
becomes a cubic equation with one real root that is given by (see
Appendix A)

\begin{widetext}

\begin{equation}
F^{10}=\frac{B_0}{2a}\Bigl(\frac{\theta}{2bB_0}\Bigr)^\frac{2}{3}\,
\biggl\{\biggl[1-\Bigl(1-\frac{32E_0^3}{27\theta^2B_0b^4}\Bigr)
^\frac{1}{2}
\biggr]^\frac{1}{3}
+\biggl[1+\Bigl(1-\frac{32E_0^3}{27\theta^2B_0b^4}\Bigr)^\frac{1}{2}
\biggr]^\frac{1}{3}\biggr\}^2
\label{4.5}
\end{equation}

\end{widetext}

\noindent
For $\theta\neq0$, the terms $32E_0^3/27\theta^2b^4B_0$ is $\ll1 $ as
$b\rightarrow\infty$. Making appropriate expansions in the above
relation, we obtain the following asymptotic and simpler solution for
$F^{10}$,

\begin{equation}
F^{10}=\frac{1}{2a}\,\Bigl(\frac{\theta^2B_0}{b^2}\Bigr)^\frac{1}{3}
\label{4.6}
\end{equation}

\noindent
With this approximation, the Einstein equations read

\begin{eqnarray}
2\,\frac{\dot a\dot b}{ab}
+\Bigl(\frac{\dot b}{b}\Bigr)^2
&=&-\,\frac{1}{2}\Bigl[\frac{B_0^2}{b^4}
+\frac{\theta}{4b}
\Bigl(\frac{\theta B_0^2}{b}\Bigr)^\frac{1}{3}\Bigr]
\nonumber\\
&&+\frac{\theta}{4\sqrt2\,b}
\Bigl(\frac{\theta B_0^2}{b}\Bigr)^\frac{1}{3}
\label{4.7}\\
2\,\frac{\ddot b}{b}
+\Bigl(\frac{\dot b}{b}\Bigr)^2
&=&-\,\frac{1}{2}\Bigl[\frac{B_0^2}{b^4}
+\frac{\theta}{4b}
\Bigl(\frac{\theta B_0^2}{b}\Bigr)^\frac{1}{3}\Bigr]
\nonumber\\
&&-\frac{\theta}{4\sqrt2\,b}
\Bigl(\frac{\theta B_0^2}{b}\Bigr)^\frac{1}{3}
\label{4.8}\\
\frac{\ddot b}{b}+\frac{\ddot a}{a}+\frac{\dot a\dot b}{ab}
&=&\frac{1}{2}\Bigl[\frac{B_0^2}{b^4}
+\frac{\theta}{4b}
\Bigl(\frac{\theta B_0^2}{b}\Bigr)^\frac{1}{3}\Bigr]
\nonumber\\
&&-\frac{\theta}{4\sqrt2\,b}
\Bigl(\frac{\theta B_0^2}{b}\Bigr)^\frac{1}{3}
\label{4.9}
\end{eqnarray}

Now, the cosmological term behaves as $b^{-\frac{4}{3}}$, that
decay slower than the previous $b^{-4}$ of the radiation term.
However, there is also a term with the behavior of $b^{-\frac{4}{3}}$
in the Maxwell counterpart and this last term avoids the obtainment of
a Friedmann behavior we are looking for.

\medskip
From this analysis, we see that the solution of (\ref{3.10}) leads to
a behavior for the cosmological term that is lower than $b^{-4}$ for
$-1<p<0$. However, there is also a term with this behavior in the
Maxwell counterpart. What may happen is that, depending on the value
of $p$, one term may dominate relatively to the other. To see if this
actually occurs we need to know the solution of (\ref{3.10}) for a
general $p$ (between zero and minus one). Of course, this is not an
easy task. However, we observe that the asymptotic solution for
$F^{10}$ given by (\ref{4.6}) could have been directly inferred from
(\ref{3.10}) by discarding the constant term $E_0$. It is then easily
seen that any asymptotic solution for any $p$ between zero and minus
one can be obtained in the same way. The result is

\begin{equation}
F^{10}=\frac{1}{2a}\biggl[-2(1+p)\theta\,
\Bigl(\frac{B_0}{b^2}\Bigr)^{1+p}\biggr]^\frac{1}{1-p}
\label{4.10}
\end{equation}

\noindent
which is in agreement with (\ref{4.6}) if one takes $p=-1/2$. We
should not worry about the minus sign inside the $\frac{1}{1-p}$-root
of equation above because $F^{10}$ always appears squared in the
calculations of significant quantities. Now, the obtainment of the
Einstein equations is just a matter of algebraic work. The result is
(we conveniently put the two terms with the same behavior for higher
$b$ together)

\begin{widetext}

\begin{eqnarray}
&&2\,\frac{\dot a\dot b}{ab}
+\Bigl(\frac{\dot b}{b}\Bigr)^2
=-\frac{B_0^2}{2b^4}
-\Bigl(\frac{p2^{-p}}{1+p}+1\Bigr)\,
\biggl[2^\frac{3p-1}{2}(1+p)\theta\,
\Bigl(\frac{B_0}{b^2}\Bigr)^{1+p}\biggr]^\frac{2}{1-p}
\label{4.11}\\
&&2\,\frac{\ddot b}{b}
+\Bigl(\frac{\dot b}{b}\Bigr)^2
=-\frac{B_0^2}{2b^4}
-\Bigl(\frac{p2^{-p}}{1+p}+1\Bigr)\,
\biggl[2^\frac{3p-1}{2}(1+p)\theta\,
\Bigl(\frac{B_0}{b^2}\Bigr)^{1+p}\biggr]^\frac{2}{1-p}
\label{4.12}\\
&&\frac{\ddot b}{b}+\frac{\ddot a}{a}+\frac{\dot a\dot b}{ab}
=\frac{B_0^2}{2b^4}
-\Bigl(\frac{p2^{-p}}{1+p}-1\Bigr)\,
\biggl[2^\frac{3p-1}{2}(1+p)\theta\,
\Bigl(\frac{B_0}{b^2}\Bigr)^{1+p}\biggr]^\frac{2}{1-p}
\label{4.13}\\
\end{eqnarray}

\end{widetext}

\noindent
where in the terms that appear $(\frac{p2^{-p}}{1+p}+1)$ or
$(\frac{p2^{-p}}{1+p}-1)$, the first part comes from the cosmological
term and the other one from the Maxwell counterpart. We then observe
that as $p$ is closed to minus one the cosmological term is more and
more dominant and, consequently, the solution tends more and more to
the Friedmann scenario. In the limit case of $p=-1$, the Maxwell term
disappears, and just remains an actual cosmological constant as it had
already been pointed out in the beginning.

\section{Conclusion}

In this paper we have analyzed further the recent
proposal\cite{Novello} of a cosmological scenario in which the
cosmological ``constant" is spacetime dependent and whose origin is
related to a primordial era, supposed dominated by quantum effects.

\medskip
In order to see if this model is actually compatible with the
observable universe, where the anisotropy rate is considerable low
[like Friedmann-Robertson-Walker (FRD)], we leave a parameter free
($p$) in the theory. We conclude that it can vary from minus one to
zero, where these limits mean an actual cosmological constant and no
cosmological term, respectively. We have show that as $p$ is closed to
minus one as the solution is compatible with the Friedmann scenario.

\medskip
A next natural step in this research line is to look for the solution
of the Einstein equations (\ref{4.11}) - (\ref{4.13}). We are
presently work in this problem and possible results shall be reported
elsewhere \cite{Novello4}.

\begin{acknowledgments}
This work is supported in part by Conselho Nacional de Desenvolvimento
Cient\'{\i}fico e Tecnol\'ogico - CNPq (Brazilian Research Agency).
One of us, J.B.-N. has also the support of PRONEX 66.2002/1998-9.
\end{acknowledgments}

\appendix
\section {Solution of (\ref{3.10}) for p=-1/2}
\renewcommand{\theequation}{A.\arabic{equation}}
\setcounter{equation}{0}

The Eq. (\ref{3.10}) for $p=-1/2$ reads

\begin{equation}
ab^2\,F^{10}+\frac{\theta b}{2}\,\sqrt\frac{B_0}{2aF^{10}}=E_0
\label{A.1}
\end{equation}

\noindent
This expression can be written in a cubic equation whose general form
is

\begin{equation}
u^3+a_1u^2+a_2u+a_3=0
\label{A.2}
\end{equation}

\noindent
where

\begin{eqnarray}
u&=&\sqrt\frac{2aF^{10}}{B_0}
\nonumber\\
a_1&=&0
\nonumber\\
a_2&=&-\,\frac{2E_0}{B_0b^2}
\nonumber\\
a_3&=&\frac{\theta}{B_0b}
\label{A.3}
\end{eqnarray}

\noindent
From the discriminant relation

\begin{equation}
D=Q^3+R^2
\label{A.4}
\end{equation}

\noindent
where

\begin{eqnarray}
&&Q=\frac{3a_2-a_1^2}{9}=-\frac{2E_0}{3B_0b^2}
\label{A.5}\\
&&R=\frac{9a_1a_2-27a_3-2a_1^3}{54}=-\frac{\theta}{2B_0b}
\label{A.6}
\end{eqnarray}

\noindent
we obtain

\begin{equation}
D=-\frac{8E_0^3}{27B_0^3b^6}+\frac{\theta^2}{4B_0^2b^2}
\label{A.7}
\end{equation}

We observe that in the region for higher $b$, $D$ is positive (for
$\theta\neq0$). Consequently, just one root is real for this region
and it
is given by

\begin{equation}
u=S+T-\frac{a_1}{3}
\label{A.8}
\end{equation}

\noindent
where

\begin{eqnarray}
S&=&\sqrt[3]{R+\sqrt D}
\nonumber\\
T&=&\sqrt[3]{R-\sqrt D}
\label{A.9}
\end{eqnarray}

\noindent
The combination of (\ref{A.6}) - (\ref{A.9}) leads to the solution
given by (\ref{4.5}).

\end{document}